\documentclass[letter,twocolumn,letterpaper]{jpsj3}

\usepackage{txfonts}

\newcommand{\figonet}{\figone}
\newcommand{\figtwot}{\figtwo}
\newcommand{\figthreet}{\figthree}
\newcommand{\figfourt}{\figfour}
\newcommand{\figones}{}
\newcommand{\figtwos}{}
\newcommand{\figthrees}{}
\newcommand{\figfours}{}

\title{High-Resolution Scanning Tunneling Spectroscopy of Vortex Cores\\ 
in Inhomogeneous Electronic States of Bi$_2$Sr$_2$CaCu$_2$O$_x$}

\author{Shunsuke YOSHIZAWA$^{1}$, 
Taiji KOSEKI$^{1}$, 
Ken MATSUBA$^{1}$, 
Takashi MOCHIKU$^{2}$,\\
Kazuto HIRATA$^{2}$, 
and Nobuhiko NISHIDA$^{1}$\thanks{Present address: Toyota Physical and Chemical Research Institute, Nagakute, Aichi 480-1192, Japan}\thanks{E-mail address: nishida@toyotariken.jp}
}

\inst{
$^{1}$Department of Physics, Tokyo Institute of Technology, Tokyo 152-8551, Japan\\
$^{2}$Superconducting Properties Unit, National Institute for Materials Science, Tsukuba, Ibaraki 305-0047, Japan
}

\abst{
We studied electronic states in vortex cores of slightly overdoped Bi$_2$Sr$_2$CaCu$_2$O$_x$ by scanning tunneling spectroscopy. 
We have found that they have stripe structures with a $4a_0$ width extending along the Cu-O bond directions.
Vortex core states are observed as two peaks at particle-hole symmetric positions in the energy gap.
Along a stripe, the peak positions of vortex core states are constant and not influenced by the spatial variation of the energy gap $\Delta$.
Outer stripes have a larger energy than inner stripes.
A mazelike pattern in the electronic states at $E = \pm\Delta$ has been observed all over the surface both inside and outside the vortex core.
The orientation of stripes of vortex core states was found to be related to the mazelike pattern in the vortex core region.
A short-range order of the mazelike pattern spatially coexists with the superconductivity and locally breaks the symmetry of the two Cu-O bond directions.
We propose that the vortex core bound states are formed by Bogoliubov quasiparticles owing to the depairing of Cooper pairs and have a local $C_2$ symmetry influenced by the short-range order of the mazelike pattern.
}

\kword{Bi2212, high-$T_c$ superconductivity, vortex core, inhomogeneity, STM, STS}

\newcommand{\figone}{
\begin{figure}
\begin{center}
\includegraphics{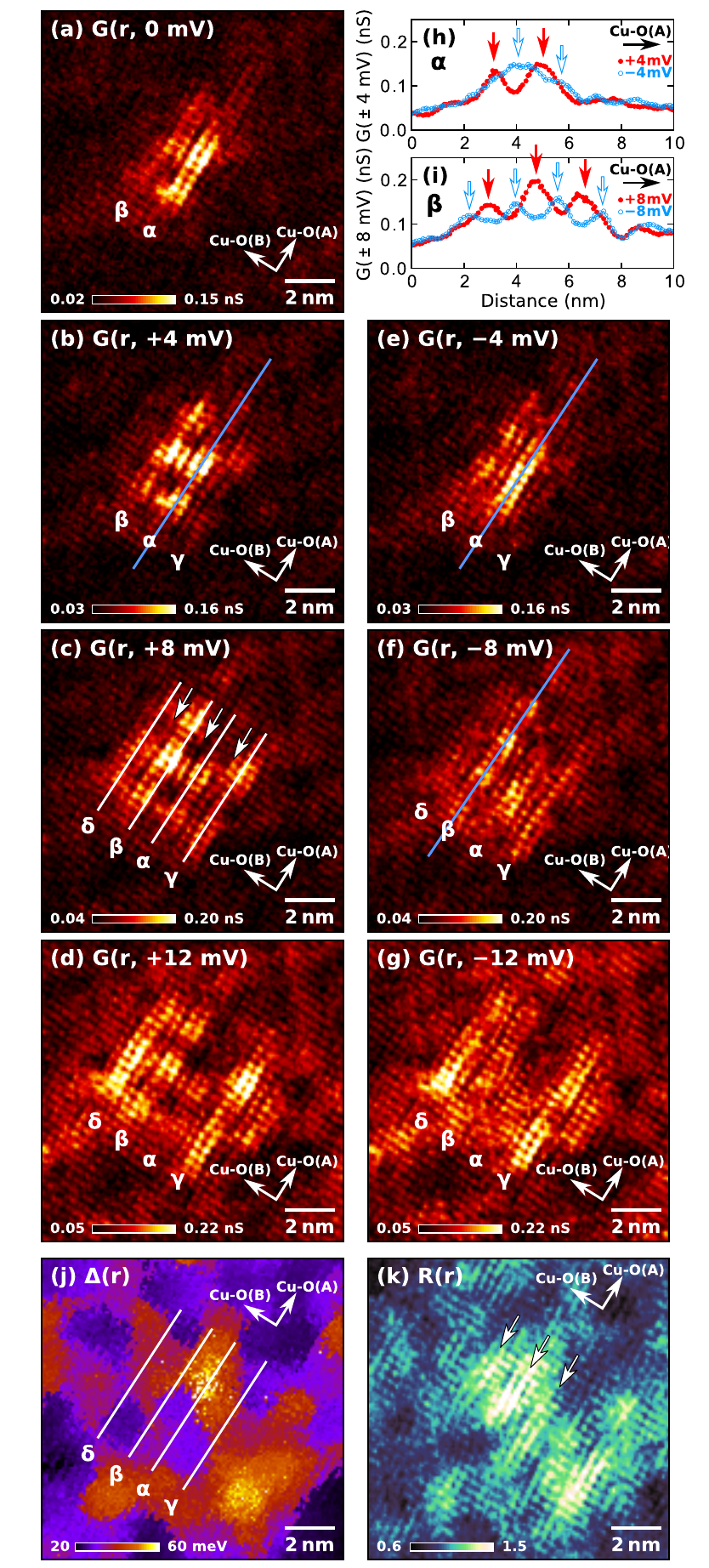}
\end{center}
\caption{(Color)
(a)-(g) Vortex core images at several bias voltages.
$G({\bf r}, V)$ of a $12 \times 12$ nm$^2$ region with a spatial resolution of 94 pm are mapped.
The tunneling conditions were $V = +200$ mV and $I = 90$ pA.
(h) Line profiles of vortex core images at $\pm 4$ mV along stripe $\alpha$.
(i) Line profiles of vortex core images at $\pm 8$ mV along stripe $\beta$.
These profiles were measured along the lines drawn in (b), (e), and (f).
(j) Energy gap $\Delta({\bf r})$ of the same region. 
The average is $\bar{\Delta} = 33$ meV.
(k) Peak height ratio $R({\bf r})$ of the same region.
}
\label{f1}
\end{figure}
}

\newcommand{\figtwo}{
\begin{figure*}
\hspace{5mm}\includegraphics[width=166mm]{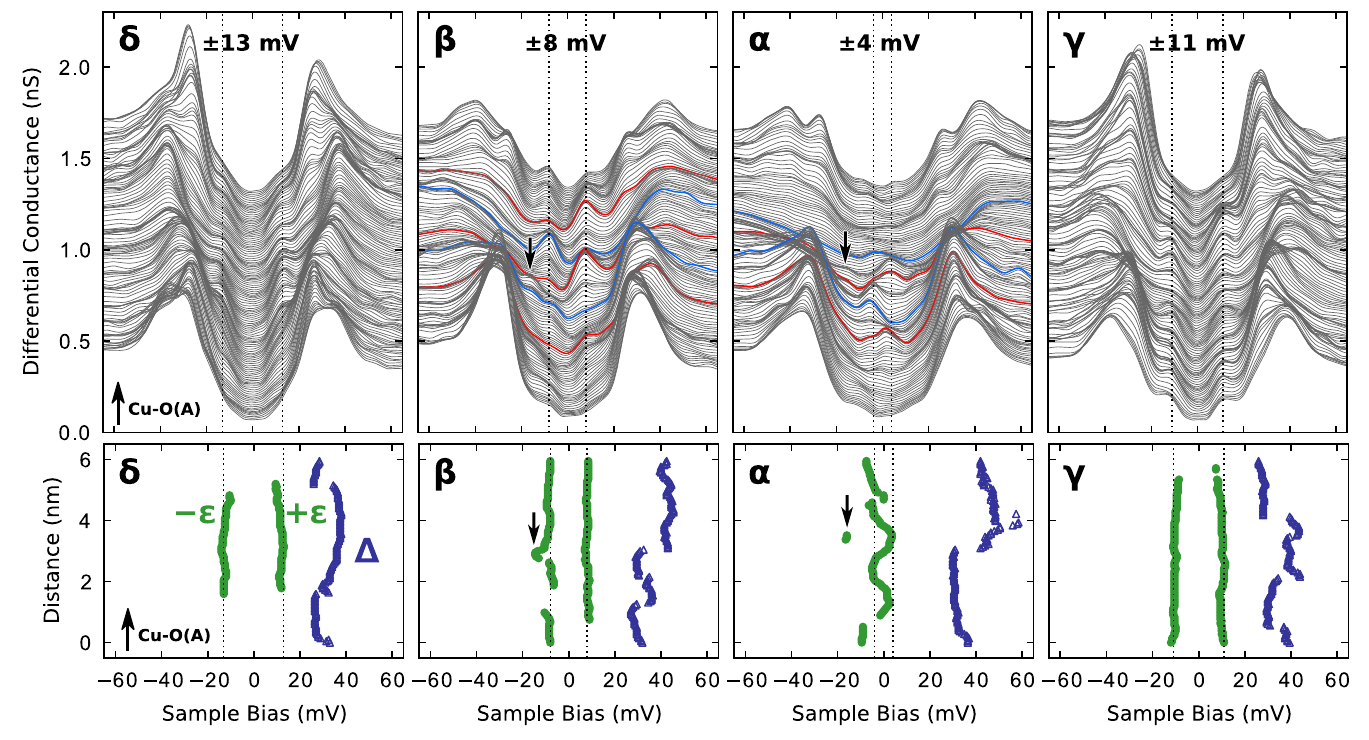}
\caption{(Color)
(Upper column) Tunneling spectra in stripes $\alpha$, $\beta$, $\gamma$, and $\delta$ measured with 47 pm spacing on white lines in the vortex core image [Fig. 1(c)] and $\Delta$ map [Fig. 1(j)].
Several curves are shown in red or blue to clarify the antiphase modulation of electron-like and hole-like states.
The tunneling conditions were $V = +200$ mV and $I = 60$ pA.
(Bottom column) Magnitude of the energy gap $\Delta$ and the peak energy of vortex core states $\varepsilon$ determined from the tunneling spectra in the upper column.
The values of $\varepsilon$ were determined from the minima in $dI^3/d^3 V$ curves.
Vertical dotted lines are guides to the eye.
In stripe $\alpha$, $\varepsilon$ exhibits an oscillating behavior, probably owing to the overlap of two peaks at about $\pm 4$ mV with antiphase amplitude oscillations.
Some of the tunneling spectra in stripes $\alpha$ and $\beta$ have a peak at about $-15$ mV, as indicated by arrows. 
This extra peak seems to be related to vortex core states, although the details are unclear.
}
\label{f2}
\end{figure*}
}

\newcommand{\figthree}{
\begin{figure}
\begin{center}
\includegraphics{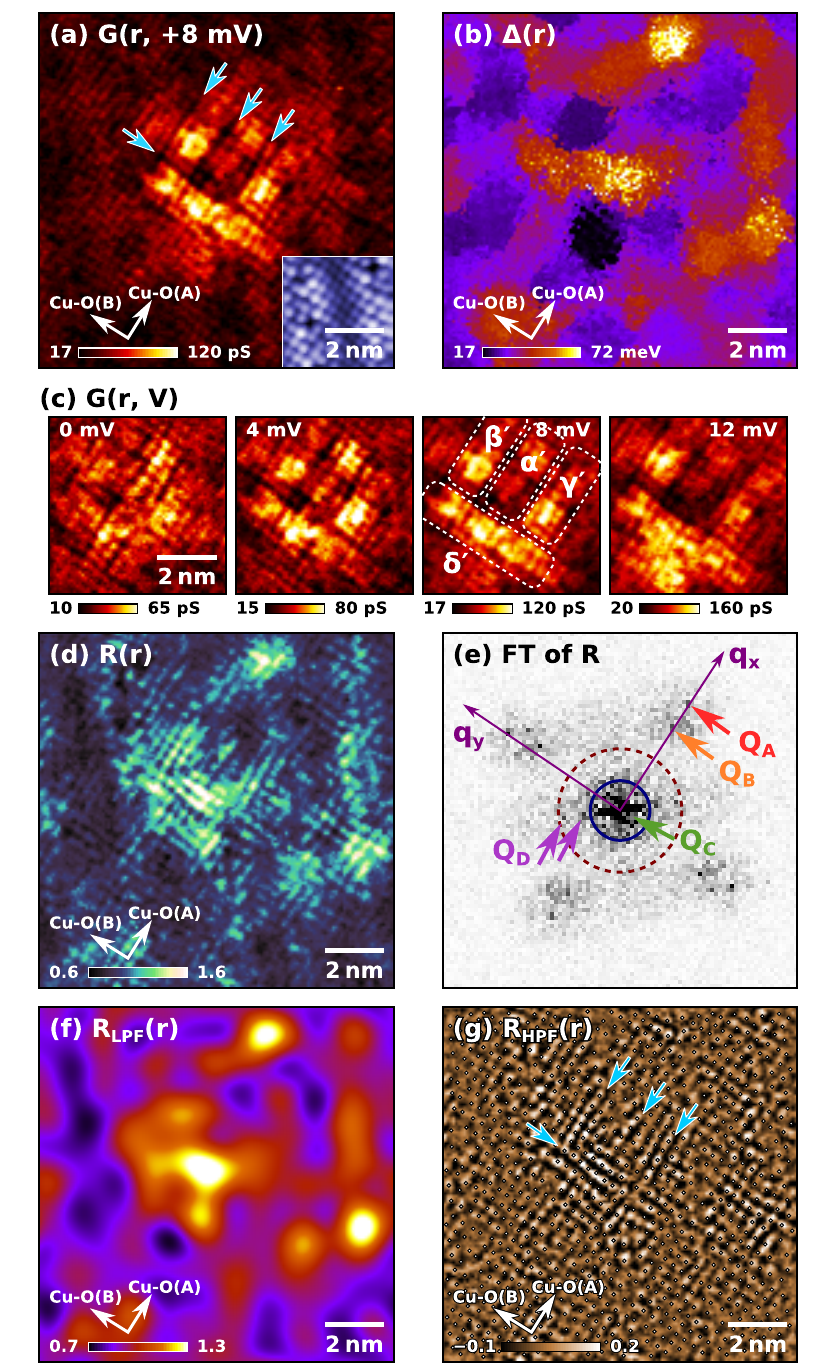}
\end{center}
\caption{(Color)
(a) Vortex core imaged by mapping $G({\bf r}, +8 {\rm mV})$ of a $12 \times 12$ nm$^2$ region with a spatial resolution of 94 pm.
The inset shows part of the STM image obtained simultaneously with the vortex core image.
The tunneling conditions were $V = +200$ mV and $I = 30$ pA.
(b) $\Delta$ map of the same region.
$\bar{\Delta}$ is 36 meV.
(c) Vortex core images at $0, +4, +8, {\rm and} +12$ mV.
(d) Peak height ratio $R({\bf r})$ of the same region as (a).
(e) Fourier transform (FT) image of the $R({\bf r})$ map.
The image shows several kinds of characteristic peaks.
The peaks $Q_A$ at $(\pm 1, 0)$ and $(0, \pm 1)$ in units of $2\pi/a_0$ represent the periodicity of the atomic lattice.
The peaks $Q_B$ at $(\pm 0.77, 0)$ and $(0, \pm 0.77)$ represent a periodicity slightly longer than $a_0$.
The broad peak near the origin ($Q_C$) represents the long-scale spatial variation.
Peaks corresponding to the supermodulation of the crystal lattice are also observed ($Q_D$).
(f) Low-pass-filtered $R({\bf r})$ map $R_{\rm LPF}({\bf r})$ showing the long-scale spatial variation related to FT peak $Q_C$.
The cutoff of the filter is 0.23 ($2\pi/a_0$) indicated by the solid circle in the FT image.
(g) High-pass-filtered $R({\bf r})$ map $R_{\rm HPF}({\bf r})$ showing the atomic-scale modulations related to the FT peaks $Q_A$ and $Q_B$.
The cutoff of the filter is 0.47 ($2\pi/a_0$), indicated by the dotted circle in the FT image.
The Cu sites determined from the STM image are marked by dots.
}
\label{f3}
\end{figure}
}

\newcommand{\figfour}{
\begin{figure}
\begin{center}
\includegraphics{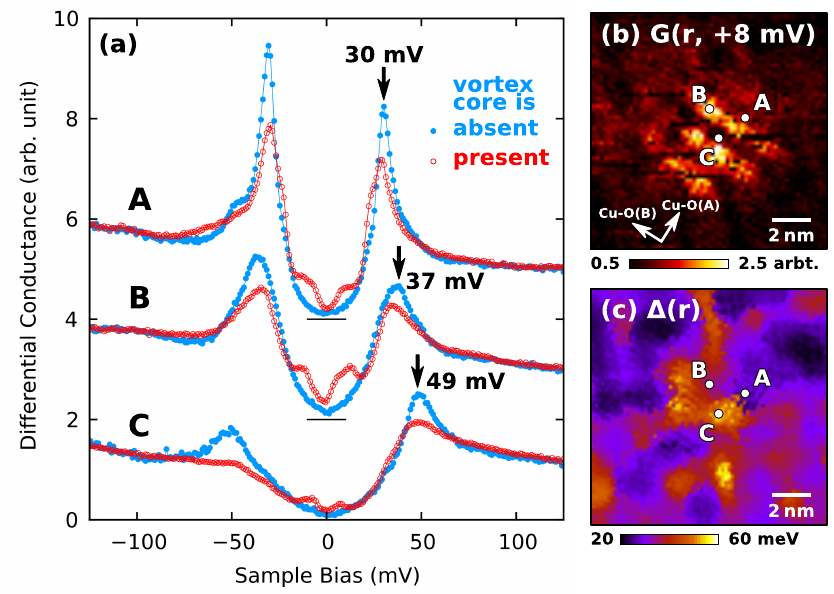}
\end{center}
\caption{(Color)
(a) Tunneling spectra measured in the presence (open circles in red) and absence (filled circles in blue) of the vortex core.
A, B, and C represent the measurement positions indicated in (b) and (c).
The spectra are offset for clarity.
The arrows indicate the $\Delta$ values in the absence of the vortex core.
The peak at $V = -\Delta/e$ is higher than that at $V = +\Delta/e$ in the spectra with a smaller $\Delta$ (A and B), whereas the opposite relation is observed in the spectra with a larger $\Delta$ (C).
This demonstrates the systematic relation between $R({\bf r})$ and $\Delta({\bf r})$ described in the main text.
A lock-in amplifier was used with a bias modulation of 2 mV$_{\rm rms}$ and 612 Hz.
The tunneling conditions were $V = +180$ mV and $I = $ 90 pA.
(b) Vortex core image at $+8$ mV.
(c) $\Delta$ map of the same region.
$\bar{\Delta}$ is 33 meV.
}
\label{f4}
\end{figure}
}

\begin{document}
\maketitle

In vortex cores of type-II superconductors, the superconducting order parameter (pair potential) is suppressed in amplitude.
Bogoliubov quasiparticles are confined in the vortex core to form Andreev bound states, reflecting the symmetry of the superconducting order parameter and the shape of the Fermi surface.
Such vortex core bound states have been observed in the conventional s-wave superconductors NbSe$_2$,\cite{Hess1990} YNi$_2$B$_2$C,\cite{Nishimori2004,Kaneko2012} and NbS$_2$,\cite{Guillamon2008} and recently in iron pnictides,\cite{Shan2011,Song2011,Hanaguri2012} by scanning tunneling spectroscopy (STS).
In d-wave superconductors, theoretical calculations based on the Bardeen-Cooper-Schrieffer theory predict that the vortex core bound states have a fourfold star-shaped spatial distribution extending in the nodal directions with a zero-bias conductance peak at the center of the vortex core.\cite{Wang1995,Schopohl1995,Ichioka1996}
However, STS experiments on the high-$T_c$ cuprate superconductor Bi$_2$Sr$_2$CaCu$_2$O$_x$ (Bi2212) showed that the vortex core has characteristic states (vortex core states) at finite excitation energies $E = \pm \varepsilon$,
and different values of $\varepsilon$ ranging from 7 to 16 meV have been reported by different research groups. \cite{Pan2000,Matsuba2003b,Hoogenboom2001}
Vortex core states exhibit spatial modulations in the two Cu-O bond directions (anti-nodal directions) with a period of about $4a_0$, where $a_0$ is the Cu-O-Cu bond length (0.38 nm).\cite{Hoffman2002,Levy2005}
Our recent study has shown that the modulation of the electron-like state and that of the hole-like state are in antiphase with each other, and detected that the modulations are commensurate ($4a_0$) in one Cu-O bond direction and incommensurate ($4.3a_0$) in the other direction.\cite{Matsuba2007}
The latter means that the vortex core states locally have the $C_2$ symmetry instead of the $C_4$ symmetry of the CuO$_2$ unit cell.
Many theoretical models\cite{Franz2002,Polkovnikov2002,Zhu2002,Tsuchiura2003,Takigawa2003,Knapp2005,Chen2005,Udby2006,Zhao2008,Tomita2009,Chen2002,Seo2008} have been proposed to explain these experimental findings, but the vortex core states in Bi2212 are far from being understood.

Vortex cores in Bi2212 are situated in inhomogeneous electronic states.
The energy gap $\Delta$, measured by STS, varies by about 3 times in magnitude on a length scale of about 2 nm.\cite{Matsuba2003a,Pan2001,Lang2002}
The vortex core is larger than the length scale of this inhomogeneity and occupies a region with dimensions of about $6 \times 6$ nm$^2$.
To understand the vortex core states, it is essential to study how they are influenced by the electronic inhomogeneity.
These studies can be performed only by STS with a high spatial resolution.
STS measurements from this point of view have not been performed so far.
Here, we report our simultaneous STS measurements of the vortex core states and the electronic inhomogeneity in Bi2212.

We used a laboratory-built cryogenic scanning tunneling microscope (STM).
A single crystalline sample of slightly overdoped Bi2212 ($T_c = $ 86 K) grown by the floating-zone method was cleaved {\it in situ} in helium atmosphere at low temperature to expose a clean surface.
A mechanically sharpened Pt-Ir wire was used as the STM tip.
Measurements were performed at 4.2 K in a magnetic field of 14.5 T applied parallel to the crystalline c-axis.
$I$-$V$ characteristics ($I$ is the tunneling current and $V$ is the sample bias voltage) were measured at each point ${\bf r}$ of an equally spaced mesh over the field of view.
They were numerically differentiated, and the differential conductance (or tunneling spectrum) $G \equiv dI/dV$ was obtained as a function of ${\bf r}$ and $V$.
Vortex cores were imaged by plotting $G$ at a low bias voltage as a function of ${\bf r}$.
The magnitude of the energy gap, $\Delta({\bf r})$, was determined from the maximum $G({\bf r}, V)$ for $V > 0$.

\figonet

The images of a vortex core are shown in Figs. 1(a)-1(g) by plotting $G({\bf r}, V)$ at several bias voltages.
The vortex core states have been found to expand in space with increasing energy.
In this vortex core, we observed four stripelike structures with a $4a_0$ width extending in the Cu-O(A) direction.
At 0 mV, bright stripe $\alpha$ and faint stripe $\beta$ are observed [Fig. 1(a)].
At +4 mV, the intensity of stripe $\beta$ increases and stripe $\gamma$ appears on the lower side of $\alpha$ [Fig. 1(b)].
At +8 mV, another stripe $\delta$ appears on the upper side of $\beta$ [Fig. 1(c)].
At +12 mV, stripe $\alpha$ becomes faint and stripes $\gamma$ and $\delta$ become distinct [Fig. 1(d)].
The images at negative biases, shown in Figs. 1(e)-1(g), exhibit similar behaviors.

The electronic states in these stripes have been studied in detail by measuring tunneling spectra with a higher spatial resolution of 47 pm, as shown in Fig. 2.
The measurement positions are marked with the lines in the vortex core image [Fig. 1(c)] and the $\Delta$ map [Fig. 1(j)].
Vortex core states are observed as two peaks at $V = \pm\varepsilon/e$ in the energy gap.
The positions of these peaks are symmetric with respect to the Fermi energy. 
In the lower column of Fig. 2, we plotted the peak energy $\varepsilon$ and the energy gap $\Delta$ as functions of position.
Along the stripes, $\Delta$ varies in space, whereas $\varepsilon$ is nearly constant.
This means that the energy of vortex core states, $\varepsilon$, is not influenced by the spatial variation of the energy gap, $\Delta$.
Our result does not support the proposal that $\varepsilon$ is proportional to $\Delta$.\cite{Hoogenboom2001}
$\varepsilon$ is smaller in inner stripes $\alpha$ (4 meV) and $\beta$ (8 meV), than in outer stripes $\gamma$ (11 meV) and $\delta$ (13 meV).

In stripes $\alpha$ and $\beta$, vortex core states exhibit spatial modulations in the Cu-O(A) direction.
In Figs. 1(h) and 1(i), we show line profiles measured along stripes $\alpha$ and $\beta$ at the peak energies in these stripes: $\pm 4$ meV for $\alpha$ and $\pm 8$ meV for $\beta$.
The spatial modulations of the electron-like state ($V > 0$) and the hole-like state ($V < 0$) are in antiphase with each other.
The periods of the modulations are determined from these profiles and are almost the same for the two stripes: ($4.5 \pm 0.3)a_0$ in stripe $\alpha$ and $(4.6 \pm 0.2)a_0$ in stripe $\beta$.
These results are consistent with the previous results.\cite{Matsuba2007}
In stripes $\gamma$ and $\delta$, such spatial modulations are not clear.

\figtwot

We discuss the nature of the vortex core states observed here.
The behavior of the vortex core states expanding in space with increasing energy is the same as that of the vortex core bound states of s-wave superconductors.\cite{Hess1990,Nishimori2004,Kaneko2012,Guillamon2008}
As the electron-like and the hole-like vortex core states are observed at symmetric positions with respect to the Fermi energy and are modulated in space in antiphase with each other, they have the characteristics of Bogoliubov quasiparticles.\cite{Zagoskin1998}
Thus, we conclude that the vortex core states in Bi2212 are the bound states of Bogoliubov quasiparticles.
The formation of the stripes breaking the $C_4$ symmetry of the CuO$_2$ suggests that some ordered state affects the structure of the bound states.
We next clarify this point.

In Fig. 1(k), we plot the ratio of the peak height at $V = +\Delta/e$ to that at $V = -\Delta/e$, $R({\bf r}) \equiv G({\bf r}, +\Delta({\bf r})/e)/G({\bf r}, -\Delta({\bf r})/e)$.
In the vortex core region, the $R({\bf r})$ map displays bright chains extending along the Cu-O(A) direction, as indicated by arrows.
We found that the direction of the bright chains is parallel to the stripes of vortex core states.
The pattern in this $R({\bf r})$ map is similar to the bond-centered pattern observed in underdoped Bi2212 in zero magnetic field.\cite{Kohsaka2007,Kohsaka2008}

In Figs. 3(a) and 3(c), another vortex core is imaged at several bias voltages.
Vortex core states expand in space with increasing energy, reproducing the result on the first vortex core.
Stripes in the two Cu-O bond directions coexist in this vortex core.
In stripe $\delta'$, $\varepsilon$ is nearly constant at $10$ meV.
In the central region $\alpha'$, the energy of vortex core states, $\varepsilon$, is distributed from 2 to 10 meV.
In regions $\beta'$ and $\gamma'$, the variation of $\varepsilon$ is within a smaller range from 8 to 12 meV.

The peak height ratio $R({\bf r})$ in this region is shown in Fig. 3(d).
The image exhibits a long-scale ($\sim$2 nm) spatial variation and atomic-scale ($\sim a_0$) spatial modulations.
The long-scale spatial variation is extracted by applying a low-pass filter to the $R({\bf r})$ map.
We found that the filtered image $R_{\rm LPF}({\bf r})$ [Fig. 3(f)] has a spatial variation similar to $\Delta({\bf r})$ [Fig. 3(b)].
This is because the peak height ratio $R$ systematically increases with increasing the magnitude of the energy gap $\Delta$.
This tendency can be seen in the tunneling spectra with different $\Delta$'s shown in Fig. 4(a) (see the figure caption).
The atomic-scale spatial modulations in the $R({\bf r})$ are extracted in the high-pass-filtered map $R_{\rm HPF}({\bf r})$ shown in Fig. 3(g).
In this image, Cu sites are linked unidirectionally, creating a mazelike complex pattern.
This indicates the presence of a short-range order locally breaking the equivalence of the two Cu-O bond directions and having an energy scale of $\Delta$.
As the mazelike pattern is observed both inside and outside the vortex core, the short-range order coexists with the superconductivity in real space.
As indicated by blue arrows in Figs. 3(a) and 3(g), the bright chains of the mazelike pattern coincide with the dark regions separating the stripes in the vortex core image.
This result means that the orientation of the stripes is affected by the short-range order.

The results on the other vortex cores are briefly mentioned here.
We studied 33 vortex cores in this experiment.
In 9 vortex cores, stripes were aligned in one Cu-O bond direction, as in Fig. 4(b).
In 6 vortex cores, stripes were found partially, or stripes in the two Cu-O bond directions were observed in a vortex core, as in Fig. 3(a).
In the other vortex cores, we could not observe clear stripes, although many of the vortex cores exhibit some kind of directionality.
The appearance of stripes can be attributed to the short-range order making the mazelike pattern. 
In almost all the vortices (29 out of 33), vortex core states were found to expand in space with increasing energy.

\figthreet

\figfourt

To clarify the nature of the electronic states at $E = \pm \Delta$, we measured the change of the electronic states caused by the presence of a vortex core as follows.
The applied magnetic field was decreased from 14.5 to 6 T with the sample temperature kept at 4.2 K.
Then, when observing a vortex core, black scars were observed in the image, as in Fig. 4(b). This is because during the measurement a vortex sometimes disappeared from the field of view and returned to the same position with atomic-scale accuracy within a time interval of the order of 1 minute.
By repeating the $dI/dV$ measurement with the STM tip fixed at one site in the vortex core, we were able to obtain the tunneling spectra in the presence and absence of the vortex core at exactly the same position.
In Fig. 4(a), we show the tunneling spectra obtained at three positions with different values of $\Delta$ [A (30 meV), B (37 meV), and C (49 meV)], marked in the vortex core image [Fig. 4(b)] and $\Delta$ map [Fig. 4(c)].
When a vortex core is present, the two peaks at $V = \pm\Delta/e$ are suppressed in intensity, and vortex core states are created at the electron-hole symmetrical position in the energy gap with the other parts of the tunneling spectra nearly unchanged.
As these vortex core states have the nature of Bogoliubov quasiparticles due to the depairing of Cooper pairs and originate from the states with the energy $\Delta$, the electronic states with the energy $\Delta$ contain the component of the one-particle excitation of a superconductor.
This means that $\Delta$ represents the magnitude of the superconducting gap.
The same measurements were performed in the other vortex cores and $\Delta$'s ranging from 20 to 55 meV were found to be the superconducting gaps in our slightly overdoped Bi2212.

In conventional s-wave superconductors, the energy gap is almost completely filled with quasiparticle bound states near the center of a vortex core.\cite{Hess1990,Nishimori2004,Kaneko2012,Guillamon2008}
In the present case of Bi2212, tunneling spectra near the center of a vortex core still exhibit a clear energy gap, as can be seen in the spectrum at position C in Fig. 4(a).
In vortex core regions, we have observed a mazelike pattern at $E = \pm\Delta$, suggesting the presence of a short-range order with an energy scale of $\Delta$.
We can conclude that the short-range order has an energy gap with the same magnitude as the superconducting gap.

Comparing the $\Delta$ maps in Figs. 1(j), 3(b), and 4(c) with the corresponding vortex core images, we have found that the vortex core centers are situated in large-$\Delta$ regions.
The experimental observation looks inconsistent with the concept of vortex pinning in conventional superconductors where vortices tend to be situated in a region of degraded superconductivity with a small $\Delta$, but a possible decrease in Cooper-pair density in a large-$\Delta$ region will explain the experimental results as follows.
If the bulk property of an increase in the average energy gap ($\bar{\Delta}$) with decreasing carrier density\cite{Hufner2008} can be applied to local quantities, large-$\Delta$ regions are expected to have a lower carrier density than small-$\Delta$ regions.
Thus, large-gap regions will have a lower density of Cooper pairs and can work as pinning centers of vortices.
This effect may be enhanced by the coexistence of the short-range order.
In Bi2212 with a larger $\bar{\Delta}$ (56 meV) than our sample, regions with $\Delta \geq 60$ meV are reported to work as pinning sites and interpreted to be non-superconducting.\cite{Fukuo2006}
This interpretation can be considered as the limit of low Cooper-pair density.

A possible picture drawn from the present studies is as follows.
A slightly overdoped Bi2212 is like a mosaic of superconductor patches with a dimension of $\sim$2 nm with different superconducting energy gaps $\Delta$.
At low temperatures, these patches are coupled with each other to form a coherent superconducting state.
Coexisting with the superconductivity in real space, there is a short-range order locally breaking the equivalence of the two Cu-O bond directions.
The short-range order has an energy gap with the same magnitude as the superconducting gap and gives rise to a mazelike pattern in the electronic states at $E = \pm\Delta$.
When a magnetic field is applied, the superconductivity is suppressed in a vortex core and bound states of the Bogoliubov quasiparticles are formed inside.
The bound states show stripe structures of $4a_0$ width with different orientations in different vortex cores, influenced by the short-range order.

\begin{acknowledgments}
This work was supported in part by MEXT KAKENHI Grant Numbers 19340096 and 2318004, and by the Global COE Program ``Nanoscience and Quantum Physics'' at Tokyo Institute of Technology.
\end{acknowledgments}


\figones
\figtwos
\figthrees
\figfours

\end{document}